# An example of a violation of the quantum inequalities for a massless scalar field in 1-1 dimensional space-time.

Dan Solomon
Rauland-Borg Corporation
Mount Prospect, IL
Email: dan.solomon@rauland.com
Oct. 31, 2010

**Abstract.**

It is well known that the energy density of a quantum state can be negative. It has been shown that there are limits on this negative energy density which are called the quantum inequalities. In this paper we will demonstrate an example of a quantum state which violates the quantum inequalities.

**1. Introduction.**

It is well known that energy density of a quantum state in some region of space can be negative [1]. However are a number of papers which show that there are limits on this phenomenon [2-4]. These limits are known as quantum inequalities. The quantum inequalities provide lower bounds on the weighted average of the energy density. They apply to systems where all external potentials are zero, i.e., free field systems. These have been investigated in a number of papers (see [5] and references, therein). It has been suggested that the lack of such limits could lead to a violation of the second law of thermodynamics [6] or lead to "exotic" phenomenon such as traversable wormholes [7].

In this paper we are interested in quantum inequalities for a massless scalar field in 1-1 dimensional space-time. For this case E. E. Flanagen [3] has shown that there are both spatial and temporal quantum inequalities that place limits on the energy density $T_{00}(x,t)$. The spatial quantum inequality is given by,

$$\int_{-\infty}^{+\infty} T_{00}(x,t)\rho(x)dx \geq \xi_{S,\min}[\rho] \qquad (1.1)$$

This is the spatial weighted average of the energy density at a certain moment in time integrated over all space. The strictly positive weighting function $\rho(x)$ satisfies,

$$\int_{-\infty}^{+\infty} \rho(v) dv = 1 \tag{1.2}$$

The temporal quantum inequality is given by,

$$\int_{-\infty}^{+\infty} T_{00}(x,t) \rho(t) dt \geq \xi_{T,\min}[\rho] \tag{1.3}$$

This is the temporal weighted average of the energy density at a given point in space over all time. Flanagan shows that,

$$\xi_{S,\min}[\rho] = \xi_{T,\min}[\rho] = -\frac{1}{24\pi} \int_{-\infty}^{+\infty} dv \frac{\rho'(v)^2}{\rho(v)} \tag{1.4}$$

where $\rho(v)$ is non-negative but can have finite support as long as it is continuous. In a region where $\rho(v)$ is zero the integrand in the above expression is replaced by zero.

In particular if we use a Lorentzian sampling function defined by,

$$\rho_L(v) = \frac{\tau}{\pi(v^2 + \tau^2)} \tag{1.5}$$

Then,

$$\xi_{S,\min}[\rho_L] = \xi_{T,\min}[\rho_L] = -\frac{1}{24\pi\tau^2} \tag{1.6}$$

The purpose of this paper is to provide an example where the above relationships are violated. In the first section of the paper we will demonstrate a violation of the spatial quantum inequality. After this we will show a violation of the temporal quantum inequality. In the last part of the paper we will show that there is a possible error in Flanagan's proof.

We start by considering an analysis of the Casimir effect by S.G. Mamaev and N.N. Trunov [8], which is also discussed in Section 1.6 of [9]. They determined the kinetic energy density of a scalar field with zero mass in 1-1 dimension space-time in the presence of a scalar potential given by,

$$V_\lambda(x) = \lambda \left[ \delta(x - a/2) + \delta(x + a/2) \right] \tag{1.7}$$

where $\lambda$ is a non-negative constant. Mamaev and Trunov show that for this system the kinetic energy density is given by,

$$T_{00,\lambda}(x) = \begin{cases} -\eta & \text{if } |x| < a/2 \\ 0 & \text{if } |x| > a/2 \end{cases} \tag{1.8}$$

where $\eta$ is a positive constant. Therefore $T_{00,\lambda}(x)$ is negative in the region between $-a/2$ and $+a/2$ and zero elsewhere.

Note that we refer to the quantity $T_{00,\lambda}(x)$ as the kinetic energy density instead of the energy density. This is because, as discussed in Section 2, the energy density includes a term that is explicitly dependent on the scalar potential. The quantity $T_{00,\lambda}(x)$, to be defined later, does not include this term. When the scalar potential is zero the kinetic energy density and energy density are given by the same expression and it is not necessary to distinguish between them.

At this point we have a system where the kinetic energy density is completely determined. The quantum inequalities are not applicable to this system because the scalar potential is not zero. This can be easily remedied by instantaneously setting the potential equal to zero. Let us suppose that at $t = 0$ the potential is removed. This is equivalent to writing the scalar potential as $V(x,t) = \theta(-t)V_\lambda(x)$ where $\theta(-t)$ is the step function. The kinetic energy density for $t < 0$ is given by (1.8). What, then, is $T_{00}(x,+\varepsilon)$ where $\varepsilon$ is an arbitrarily small positive number? As will be discussed below an abrupt change in the potential at $t = 0$ will not cause an abrupt change in $T_{00}(x,t)$ in the vicinity of $t = 0$. $T_{00}(x,t)$ will be continuous across the boundary $t = 0$ which means that for $\varepsilon \to 0$, $T_{00}(x,+\varepsilon) = T_{00}(x,-\varepsilon) = T_{00,\lambda}(x)$. The result is that we now have a free field system in which the kinetic energy density is given by (1.8) at time $t = \varepsilon \to 0$. Since the energy density is equivalent to the kinetic energy density when the scalar potential is zero then, at time $t = \varepsilon \to 0$, the energy density is known and is equal to $T_{00,\lambda}(x)$. It is easy to show that this energy density violates the spatial quantum inequality. For example let the weighting function be the Lorentzian as defined in Eq. (1.5) then,

$$\int_{-\infty}^{+\infty} T_{00,\lambda}(x)\rho_L(x)dx = -\eta \int_{-a/2}^{+a/2} \rho_L(x)dx = -\frac{2\eta}{\pi}\arctan\left(\frac{a}{2\tau}\right) \tag{1.9}$$

If the spatial quantum inequality is valid then we can use this result along with (1.6) in (1.3) to obtain,

$$-\frac{2\eta}{\pi} arctan\left(\frac{a}{2\tau}\right) \geq -\frac{1}{24\pi\tau^2} \tag{1.10}$$

In the limit that $\tau >> a$, so that $a/2\tau$ is small, use $arctan(a/2\tau) \cong a/2\tau$ in the above to obtain,

$$-\frac{2\eta}{\pi}\frac{a}{2\tau} \underset{\tau>>a}{\geq} -\frac{1}{24\pi\tau^2} \tag{1.11}$$

From this we obtain $-\eta a \underset{\tau>>a}{\geq} -1/24\tau$ which yields $1/24\tau \underset{\tau>>a}{\geq} \eta a$. This last equality is not true for sufficiently large $\tau$. As $\tau \to \infty$, Eq. (1.11) becomes $0 \underset{\tau\to\infty}{\geq} \eta a$ which is obviously false because the right hand side is positive. Therefore the spatial quantum inequality is violated in this situation.

## 2. Continuity of the Kinetic Energy.

A key claim in this paper is that kinetic energy density is continuous with respect to an instantaneous change in the scalar potential. In this section we will show why this claim is reasonable. A more detailed analysis will be given in the next Sections.

To show why this claim makes sense we will consider a couple of simple examples in 1-1 dimensional space-time. First consider a classical particle of mass $m$. The kinetic energy of the particle is given by,

$$KE(t) = \frac{1}{2}m\left(\frac{dx}{dt}\right)^2 \tag{2.1}$$

The total energy of the particle of is,

$$E = KE + V(x,t) \tag{2.2}$$

where $V(x,t)$ is the potential. The equation of motion is given by,

$$m\frac{d^2x}{dt^2} = -\frac{\partial V(x,t)}{\partial x} \tag{2.3}$$

Let $V(x,t)$ be given by,

$$V(x,t) = \theta(-t)V(x) \tag{2.4}$$

In this case Eq. (2.3) becomes,

$$m\frac{d^2x}{dt^2} = -\frac{\partial V(x,t)}{\partial x} \text{ for } t \leq 0 \text{ and } m\frac{d^2x}{dt^2} = 0 \text{ for } t > 0 \tag{2.5}$$

The boundary conditions at $t = 0$ are,

$$x(\varepsilon) \underset{\varepsilon \to 0}{=} x(-\varepsilon) \text{ and } \frac{dx(\varepsilon)}{dt} \underset{\varepsilon \to 0}{=} \frac{dx(-\varepsilon)}{dt} \tag{2.6}$$

Using this and the definition of the kinetic energy in Eq. (2.1) it is evident that $KE(\varepsilon) \underset{\varepsilon \to 0}{=} KE(-\varepsilon)$. Therefore the kinetic energy of a particle is continuous across the boundary at $t = 0$.

For our second example consider the classical (non-quantized) Klein-Gordon equation for a zero mass particle in the presence of a scalar potential. In this case the scalar field $\varphi(x,t)$ obeys the equation by,

$$\frac{\partial^2 \varphi(x,t)}{\partial t^2} - \frac{\partial^2 \varphi(x,t)}{\partial x^2} + V(x,t)\varphi(x,t) = 0 \tag{2.7}$$

The kinetic energy density is given by,

$$T_{00}(x,t) = \frac{1}{2}\left(\frac{\partial \varphi}{\partial t} \cdot \frac{\partial \varphi}{\partial t} + \frac{\partial \varphi}{\partial x} \cdot \frac{\partial \varphi}{\partial x}\right) \tag{2.8}$$

The energy density is given by,

$$E(x,t) = T_{00} + \frac{1}{2}V\varphi^2 \tag{2.9}$$

Note that the energy density is explicitly dependent on the scalar potential. When the scalar potential is zero the energy density and kinetic energy density are equivalent expressions.

Assume the scalar potential is given by (2.4). In this case (2.7) becomes,

$$\frac{\partial^2 \varphi(x,t)}{\partial t^2} - \frac{\partial^2 \varphi(x,t)}{\partial x^2} + V(x)\varphi(x,t) = 0 \text{ for } t \leq 0 \tag{2.10}$$

and,

$$\frac{\partial^2 \varphi(x,t)}{\partial t^2} - \frac{\partial^2 \varphi(x,t)}{\partial x^2} = 0 \text{ for } t > 0 \tag{2.11}$$

The boundary conditions at $t = 0$ are given by,

$$\varphi(x,\varepsilon) \underset{\varepsilon \to 0}{=} \varphi(x,-\varepsilon) \text{ and } d\varphi(x,\varepsilon)/dt \underset{\varepsilon \to 0}{=} d\varphi(x,-\varepsilon)/dt \tag{2.12}$$

Using these boundary conditions and (2.8) it is evident that,

$$T_{00}(x,\varepsilon) \underset{\varepsilon \to 0}{=} T_{00}(x,-\varepsilon) \tag{2.13}$$

Therefore kinetic energy density of a classical scalar field is continuous across the boundary at $t=0$.

In the following discussion we will show that these results apply to a quantized scalar field and show that the kinetic energy density of the quantized scalar field is also continuous during an instantaneous removal of the scalar potential. As discussed in the introduction, this will result in a violation of the spatial quantum inequality.

**3. Mamaev and Trunov's solution.**

In this section we will derive the field operator for the system in question. We will work in the Heisenberg picture. In this case the state vector $|\Omega\rangle$ is constant in time and the time dependence of the system is given by the equations of motion of the field operator $\hat{\varphi}(x,t)$. For a zero mass scalar field in 1-1 dimensions this is given by,

$$\frac{\partial^2 \hat{\varphi}(x,t)}{\partial t^2} - \frac{\partial^2 \hat{\varphi}(x,t)}{\partial x^2} + V(x,t)\hat{\varphi}(x,t) = 0 \tag{3.1}$$

where $V(x,t)$ is the scalar potential. This is the same as Eq. (2.7) for the scalar field with $\hat{\varphi}(x,t)$ replaced by $\varphi(x,t)$.

For the specific problem to be considered here the field operator will be designated by $\hat{\varphi}_\lambda(x,t)$. This field operator is given by,

$$\hat{\varphi}_\lambda(x,t) = \sum_{j\omega}\left(\hat{a}_{j\omega}f_{\lambda,j\omega}(x,t) + \hat{a}^*_{j\omega}f^*_{\lambda,j\omega}(x,t)\right) \tag{3.2}$$

where $\hat{a}_{j\omega}$ and $\hat{a}^*_{j\omega}$ are the destruction and creation operators, respectively. They satisfy the usual commutation relationships $\left[\hat{a}_{j\omega},\hat{a}^*_{j'\omega'}\right] = \delta_{j'j}\delta_{\omega'\omega}$ with all other commutations being zero. The modes $f_{\lambda,j\omega}(x,t)$ are solutions of (3.1) with $V(x,t)$ given by,

$$V(x,t) = \begin{cases} V_\lambda(x) \text{ for } t<0 \\ 0 \text{ for } t \geq 0 \end{cases} \tag{3.3}$$

where $V_\lambda(x)$ is define by Eq. (1.7). Therefore the $f_{\lambda,j\omega}(x,t)$ are given by,

$$\frac{\partial^2 f_{\lambda,j\omega}(x,t)}{\partial t^2} - \frac{\partial^2 f_{\lambda,j\omega}(x,t)}{\partial x^2} + V_\lambda(x) f_{\lambda,j\omega}(x,t) = 0 \text{ for } t < 0 \tag{3.4}$$

and,

$$\frac{\partial^2 f_{\lambda,j\omega}(x,t)}{\partial t^2} - \frac{\partial^2 f_{\lambda,j\omega}(x,t)}{\partial x^2} = 0 \text{ for } t \geq 0 \tag{3.5}$$

The solutions to (3.4) have by been given by Mamaev and Trunov [8]. For this case $f_{\lambda,j\omega}(x,t)$ are given by,

$$f_{\lambda,j\omega}(x,t)\Big|_{t<0} = \frac{e^{-i\omega t}}{\sqrt{2\pi\omega}} \chi_{\lambda,j\omega}(x) \tag{3.6}$$

where the $\chi_{\lambda,j\omega}(x)$ are the eigensolutions to the equation,

$$-\omega^2 \chi_{\lambda,j\omega} - \frac{\partial^2 \chi_{\lambda,j\omega}}{\partial x^2} + V_\lambda(x) \chi_{\lambda,j\omega} = 0 \tag{3.7}$$

There are two sets of solutions that are designated by $j=1$ and $j=2$, respectively, and $\omega$ takes on all values from 0 to $\infty$. These solutions are given in the Appendix. Note that the $\chi_{\lambda,j\omega}(x)$ are real.

Next we will find the solutions for $f_{\lambda,j\omega}(x,t)$ for $t \geq 0$. When $t \geq 0$ we will represent $f_{\lambda,j\omega}(x,t)$ by $f^{(+)}_{\lambda,j\omega}(x,t)$. The solutions to (3.5) can be written as,

$$f^{(+)}_{\lambda,j\omega}(x,t) = C_{1j\omega}(x-t) + C_{2j\omega}(x+t) \tag{3.8}$$

where the functions $C_{1j\omega}$ and $C_{2j\omega}$ will be determined in the following discussion.

The boundary conditions at $t=0$ are given by assuming that $f_{\lambda,j\omega}(x,t)$ and its first derivative with respect to time are continuous. This yields,

$$f^{(+)}_{\lambda,j\omega}(x,0) = f_{\lambda,j\omega}(x,0) \text{ and } \partial f^{(+)}_{\lambda,j\omega}(x,0)/\partial t = \partial f_{\lambda,j\omega}(x,0)/\partial t \tag{3.9}$$

Using (3.6) in the above we obtain,

$$f^{(+)}_{\lambda,j\omega}(x,0) = \frac{\chi_{\lambda,j\omega}(x)}{\sqrt{2\pi\omega}} \text{ and } \frac{\partial f^{(+)}_{\lambda,j\omega}(x,0)}{\partial t} = \frac{-i\omega \chi_{\lambda,j\omega}(x)}{\sqrt{2\pi\omega}} \tag{3.10}$$

Use this result along with (3.8) to obtain,

$$\frac{\chi_{\lambda,j\omega}(x)}{\sqrt{2\pi\omega}} = C_{1j\omega}(x) + C_{2j\omega}(x) \tag{3.11}$$

and,

$$\frac{-i\omega \chi_{\lambda,j\omega}(x)}{\sqrt{2\pi\omega}} = -\frac{\partial C_{1j\omega}(x)}{\partial x} + \frac{\partial C_{2j\omega}(x)}{\partial x} \tag{3.12}$$

where we have used $\partial C_{1j\omega}(x-t)/\partial t = -\partial C_{1j\omega}(x-t)/\partial x$.

Next, differentiating (3.11) with respect to $x$ we obtain,

$$\frac{1}{\sqrt{2\pi\omega}} \frac{\partial \chi_{\lambda,j\omega}(x)}{\partial x} = \frac{\partial C_{1j\omega}(x)}{\partial x} + \frac{\partial C_{2j\omega}(x)}{\partial x} \tag{3.13}$$

Use (3.12) and (3.13) to obtain,

$$\frac{\partial C_{2j\omega}(x)}{\partial x} = \frac{1}{2\sqrt{2\pi\omega}} \left( \frac{\partial \chi_{\lambda,j\omega}(x)}{\partial x} - i\omega \chi_{\lambda,j\omega}(x) \right) \tag{3.14}$$

and,

$$\frac{\partial C_{1j\omega}(x)}{\partial x} = \frac{1}{2\sqrt{2\pi\omega}} \left( \frac{\partial \chi_{\lambda,j\omega}(x)}{\partial x} + i\omega \chi_{\lambda,j\omega}(x) \right) \tag{3.15}$$

Integrate the above to obtain,

$$C_{2j\omega}(x) = \frac{1}{2\sqrt{2\pi\omega}} \left( \chi_{\lambda,j\omega}(x) - i\omega \int_0^x \chi_{\lambda,j\omega}(y) dy \right) \tag{3.16}$$

and,

$$C_{1j\omega}(x) = \frac{1}{2\sqrt{2\pi\omega}} \left( \chi_{\lambda,j\omega}(x) + i\omega \int_0^x \chi_{\lambda,j\omega}(y) dy \right) \tag{3.17}$$

Use these results in (3.8) to yield,

$$f^{(+)}_{\lambda,j\omega}(x,t) = \frac{1}{2\sqrt{2\pi\omega}} \left( \chi_{\lambda,j\omega}(x-t) + \chi_{\lambda,j\omega}(x+t) - i\omega \int_{x-t}^{x+t} \chi_{\lambda,j\omega}(y) dy \right) \tag{3.18}$$

When we use the solutions for $\chi_{\lambda,j\omega}$ given in the Appendix we have the following interesting result. Consider the region of space where $|x|-t > a/2$ with $t \geq 0$. In this region $f^{(+)}_{\lambda,j\omega}(x,t) = f_{\lambda,j\omega}(x,t)$. The reason for this is that the removal of the potential at time $t=0$ creates a disturbance which moves out from the region $|x| < a/2$ at the speed of light. The portion of the mode solution that is outside the light cone, i.e. the region where $|x|-t > a/2$, does not yet "know" that the potential has been removed.

## 4. Energy density.

In this section we will solve for the kinetic energy density and show that the kinetic energy density is continuous across the boundary at $t=0$ which will confirm the results in Section 1. The kinetic energy density operator is defined by,

$$\hat{T}_{00}[\hat{\varphi}] = \frac{1}{2}\left(\frac{\partial \hat{\varphi}}{\partial t} \cdot \frac{\partial \hat{\varphi}}{\partial t} + \frac{\partial \hat{\varphi}}{\partial x} \cdot \frac{\partial \hat{\varphi}}{\partial x}\right) \quad (4.1)$$

Now that the kinetic energy density operator and field operator have been defined the next step in order to completely specify the system is to specify the time independent state vector on which the field operator acts. The state vector will be designated by $|0\rangle$ and is defined by the relationship $\hat{a}_{j\omega}|0\rangle = 0$. That is, it is destroyed by all destruction operators.

The kinetic energy density expectation value is, then, given by,

$$T_{00,\lambda} = \langle 0|\hat{T}_{00}[\hat{\varphi}_\lambda]|0\rangle \quad (4.2)$$

Now there is a problem with this expression. It can easily be shown to be infinite. However we are not really interested in absolute magnitude of the energy density but of the difference between this kinetic energy density and the kinetic energy density of the unperturbed vacuum state. Therefore we define the regularized kinetic energy density by the expression,

$$\Delta T_{00,\lambda} = \langle 0|\hat{T}_{00}[\hat{\varphi}_\lambda]|0\rangle - \langle 0|\hat{T}_{00}[\hat{\varphi}_0]|0\rangle \quad (4.3)$$

where $\hat{\varphi}_0$ is "free field" solution of the field operator for the case where the scalar potential is zero. It is obtained by setting $\lambda = 0$ in Eqs. (3.2) through (3.7). The mode solutions for this case are also given in the Appendix.

There is still a problem with evaluating this expression due to the fact that we are subtracting one infinity from another which has potential problems. Mamaev and Trunov [8] resolved this problem by using mode regularization. That is they determined the change in the kinetic energy density of each mode and then added up all the changes. Therefore the regularized kinetic energy density may be written as,

$$\Delta T_{00,\lambda}(x,t) = \sum_{j\omega}\left(\varepsilon_{\lambda,j\omega}(x,t) - \varepsilon_{0,j\omega}(x)\right) \quad (4.4)$$

where,

$$\varepsilon_{\lambda,j\omega}(x,t) = \frac{1}{2}\left[\left|\frac{\partial f_{\lambda,j\omega}(x,t)}{\partial t}\right|^2 + \left|\frac{\partial f_{\lambda,j\omega}(x,t)}{\partial x}\right|^2\right] \quad (4.5)$$

The quantity $\varepsilon_{\lambda,j\omega}(x,t)$ is the kinetic energy density of the mode "$j\omega$". The quantity $\varepsilon_{0,j\omega}(x)$ is given by the above expression with $f_{\lambda,j\omega}$ replaced by $f_{0,j\omega}$. Note that $\varepsilon_{0,j\omega}(x)$ is a constant independent of $x$ and in the following will be written as $\varepsilon_{0,j\omega}$.

For $t<0$ we use (3.6) to obtain,

$$\varepsilon_{\lambda,j\omega}(x)\Big|_{t<0} = \frac{1}{2(2\pi\omega)}\left[\omega^2\left(\chi_{\lambda,j\omega}(x)\right)^2 + \left(\frac{\partial \chi_{\lambda,j\omega}(x)}{\partial x}\right)^2\right] \quad (4.6)$$

where $\varepsilon_{\lambda,j\omega}(x)$ is time independent for $t<0$. Using the above results we obtain,

$$\Delta T_{00,\lambda}(x)\Big|_{t<0} = \sum_{j\omega}\left(\varepsilon_{\lambda,j\omega}(x) - \varepsilon_{0,j\omega}\right) \quad (4.7)$$

Mamaev and Trunov[8] have calculated this kinetic energy density as,

$$\Delta T_{00,\lambda}(x) = \begin{cases} -\eta = \eta_1 + \eta_2 \text{ if } |x| < a/2 \\ 0 \text{ if } |x| > a/2 \end{cases} \quad (4.8)$$

which can also be written as,

$$\Delta T_{00,\lambda}(x) = -\eta\left[\theta(x+a/2) - \theta(x-a/2)\right] \quad (4.9)$$

where,

$$\eta_1 = -\frac{\Lambda}{\pi a^2}\int_0^\infty \frac{ye^{-y}dy}{ye^y + \Lambda\sinh y}; \quad \eta_2 = \frac{\Lambda}{\pi a^2}\int_0^\infty \frac{ye^{-y}dy}{ye^y + \Lambda\cosh y} \quad (4.10)$$

with $\Lambda = \lambda a/2$. It can be shown that $\eta_1 + \eta_2 < 0$ based on the fact that $\cosh y \geq \sinh y$. Therefore in the region between $-a/2$ and $+a/2$ the kinetic energy density is negative and independent of $x$ within this region. Outside of this region the kinetic energy density is zero.

Next consider the kinetic energy density for $t \geq 0$, after the potential has been removed. In this case the kinetic energy density is equivalent to the energy density because the scalar potential is zero. Therefore we will refer to this quantity simply as the energy density. For $t \geq 0$ we represent the energy density of a given mode "$j\omega$" by $\varepsilon^{(+)}_{\lambda,j\omega}(x,t)$ where,

$$\varepsilon^{(+)}_{\lambda,j\omega}(x,t) = \frac{1}{2}\left[\left|\frac{\partial f^{(+)}_{\lambda,j\omega}(x,t)}{\partial t}\right|^2 + \left|\frac{\partial f^{(+)}_{\lambda,j\omega}(x,t)}{\partial x}\right|^2\right] \quad (4.11)$$

From (3.18) we obtain,

$$\frac{\partial f^{(+)}_{\lambda,j\omega}(x,t)}{\partial t} = \frac{1}{2\sqrt{2\pi\omega}}\left[\left(-\frac{\partial \chi_{\lambda,j\omega}(x-t)}{\partial x} + \frac{\partial \chi_{\lambda,j\omega}(x+t)}{\partial x}\right) - i\omega\left(\chi_{\lambda,j\omega}(x+t) + \chi_{\lambda,j\omega}(x-t)\right)\right]$$

(4.12)

and,

$$\frac{\partial f^{(+)}_{\lambda,j\omega}(x,t)}{\partial x} = \frac{1}{2\sqrt{2\pi\omega}}\left[\left(\frac{\partial \chi_{\lambda,j\omega}(x-t)}{\partial x} + \frac{\partial \chi_{\lambda,j\omega}(x+t)}{\partial x}\right) - i\omega\left(\chi_{\lambda,j\omega}(x+t) - \chi_{\lambda,j\omega}(x-t)\right)\right]$$

(4.13)

Use the fact that $\chi_{\lambda,j\omega}$ is real to obtain,

$$\left|\frac{\partial f^{(+)}_{\lambda,j\omega}(x,t)}{\partial t}\right|^2 = \frac{1}{8\pi\omega}\left[\left(-\frac{\partial \chi_{\lambda,j\omega}(x-t)}{\partial x} + \frac{\partial \chi_{\lambda,j\omega}(x+t)}{\partial x}\right)^2 + \omega^2\left(\chi_{\lambda,j\omega}(x+t) + \chi_{\lambda,j\omega}(x-t)\right)^2\right]$$

(4.14)

and,

$$\left|\frac{\partial f^{(+)}_{\lambda,j\omega}(x,t)}{\partial x}\right|^2 = \frac{1}{8\pi\omega}\left[\left(\frac{\partial \chi_{\lambda,j\omega}(x-t)}{\partial x} + \frac{\partial \chi_{\lambda,j\omega}(x+t)}{\partial x}\right)^2 + \omega^2\left(\chi_{\lambda,j\omega}(x+t) - \chi_{\lambda,j\omega}(x-t)\right)^2\right]$$

(4.15)

Use the above in (4.11) to obtain,

$$\varepsilon^{(+)}_{\lambda,j\omega}(x,t) = \frac{1}{8\pi\omega}\left\{\left[\left(\frac{\partial \chi_{\lambda,j\omega}(x-t)}{\partial x}\right)^2 + \left(\frac{\partial \chi_{\lambda,j\omega}(x+t)}{\partial x}\right)^2\right] + \omega^2\left[\chi^2_{\lambda,j\omega}(x+t) + \chi^2_{\lambda,j\omega}(x-t)\right]\right\}$$

(4.16)

Using (4.6) this can be written as,

$$\varepsilon_{\lambda,j\omega}^{(+)}(x,t) \underset{t\geq 0}{=} \frac{1}{2}\left(\varepsilon_{\lambda,j\omega}(x+t)+\varepsilon_{\lambda,j\omega}(x-t)\right) \tag{4.17}$$

Let the energy density for $t \geq 0$ be designated by $\Delta T_{00,\lambda}^{(+)}(x,t)$. Therefore we can write,

$$\Delta T_{00,\lambda}^{(+)}(x,t) = \sum_{j\omega}\left(\varepsilon_{\lambda,j\omega}^{(+)}(x,t) - \varepsilon_{0,j\omega}\right) \tag{4.18}$$

Use (4.17) in the above to obtain,

$$\Delta T_{00,\lambda}^{(+)}(x,t) = \sum_{j\omega}\left[\frac{1}{2}\left(\varepsilon_{\lambda,j\omega}(x+t)-\varepsilon_{0,j\omega}\right)+\frac{1}{2}\left(\varepsilon_{\lambda,j\omega}(x-t)-\varepsilon_{0,j\omega}\right)\right] \tag{4.19}$$

Refer to (4.7) obtain,

$$\Delta T_{00,\lambda}^{(+)}(x,t) \underset{t\geq 0}{=} \frac{1}{2}\left(\Delta T_{00,\lambda}(x-t)+\Delta T_{00,\lambda}(x+t)\right) \tag{4.20}$$

Recall that $\Delta T_{00,\lambda}(x)$ is given by (4.9). Use this in (4.20) to obtain,

$$\Delta T_{00,\lambda}^{(+)}(x,t) = \frac{-\eta}{2}\left\{\begin{array}{l}\left[\theta(x-t+a/2)-\theta(x-t-a/2)\right] \\ +\left[\theta(x+t+a/2)-\theta(x+t-a/2)\right]\end{array}\right\} \tag{4.21}$$

At $t=0$ this becomes $\Delta T_{00,\lambda}^{(+)}(x,0) = -\eta\left[\theta(x+a/2)-\theta(x-a/2)\right]$. Therefore $\Delta T_{00,\lambda}^{(+)}(x,0) = \Delta T_{00,\lambda}(x)$ which verifies that the kinetic energy density is continuous across the boundary at $t=0$. This confirms the results of Section 1 and shows that the spatial quantum inequality is violated.

Another way to look at this problem is to make use of the fact the instantaneous removal of the scalar potential does not change the total kinetic energy. The kinetic energy that is present is merely redistributed. That is it moves out from the confined area at the speed of light. To understand why this is consider the kinetic energy associated with a single mode. The kinetic energy density for a given mode is given by (4.5). The total kinetic energy associated with this mode is then,

$$E_{\lambda,j\omega}(t) = \int \varepsilon_{\lambda,j\omega}(x,t)dx = \frac{1}{2}\int\left[\left|\frac{\partial f_{\lambda,j\omega}(x,t)}{\partial t}\right|^2 + \left|\frac{\partial f_{\lambda,j\omega}(x,t)}{\partial x}\right|^2\right]dx \tag{4.22}$$

Take the time derivative of the above and note that the $f_{\lambda,j\omega}(x,t)$ satisfies (3.1) to obtain,

$$\frac{\partial E_{\lambda,j\omega}(t)}{\partial t} = \frac{1}{2}\int\left[\begin{array}{l}\left(\dfrac{\partial^2 f^*_{\lambda,j\omega}}{\partial x^2}-Vf^*_{\lambda,j\omega}\right)\dfrac{\partial f_{\lambda,j\omega}}{\partial t}+\dfrac{\partial f^*_{\lambda,j\omega}}{\partial t}\left(\dfrac{\partial^2 f_{\lambda,j\omega}}{\partial x^2}-Vf_{\lambda,j\omega}\right)\\ +\dfrac{\partial^2 f^*_{\lambda,j\omega}}{\partial t\partial x}\dfrac{\partial f_{\lambda,j\omega}}{\partial x}+\dfrac{\partial f^*_{\lambda,j\omega}}{\partial x}\dfrac{\partial^2 f_{\lambda,j\omega}}{\partial t\partial x}\end{array}\right]dx \quad (4.23)$$

Integrate by parts and rearrange terms to obtain,

$$\frac{\partial E_{\lambda,j\omega}(t)}{\partial t}=-\frac{1}{2}\int V(x,t)\frac{\partial\left|f_{\lambda,j\omega}(x,t)\right|^2}{\partial t}dx \quad (4.24)$$

Next use (3.3) to obtain,

$$\frac{\partial E_{\lambda,j\omega}(t)}{\partial t}=\left\{\begin{array}{l}-\dfrac{1}{2}\int V_\lambda(x)\dfrac{\partial\left|f_{\lambda,j\omega}(x,t)\right|^2}{\partial t}dx\text{ for }t<0\\ 0\text{ for }t\geq 0\end{array}\right. \quad (4.25)$$

For $t<0$ the $f_{\lambda,j\omega}(x,t)$ are given by (3.6). It is evident from this that, for $t<0$, the quantity $\left|f_{\lambda,j\omega}(x,t)\right|^2$ is time independent. Therefore $\partial\left|f_{\lambda,j\omega}(x,t)\right|^2/\partial t=0$ in the above expression. The result is that $\partial E_{\lambda,j\omega}(t)/\partial t=0$ for all time for the problem that we are considering. This confirms the previous result that the kinetic energy density is continuous across the boundary at $t=0$. The total kinetic energy of each mode is a constant independent of time. When the potential is removed at $t=0$ the total kinetic energy does not change but the location of the kinetic energy, that is the kinetic energy density, can change. However this "rearrangement" of energy can only occur at the speed of light.

    This is somewhat analogous to what happens to the electromagnetic field when a constant current source is turned off. Consider a constant current moving in a wire loop. There is a static magnetic field surrounding the wire. There is energy associated with this magnetic field. If the current is abruptly turned off the energy density of the electromagnetic field does not instantaneously change. After the current is turned off a radiated field is produced that will move out at the speed of light. The energy contained in this radiated field is the same as what was in the original static magnetic field. That is, the action of instantaneously turning off the current does not change the amount of energy that was in the electromagnetic field.

## 5. Quantum interest conjecture.

According to (4.21) the energy density for $t \geq 0$ consists of a square pulse of negative energy moving along the positive direction and an equivalent negative energy pulse moving in the negative direction. Both pulses move at the speed of light.

This violates the quantum interest conjecture that was originally proposed by Ford and Roman [10]. According to this conjecture any pulse of negative energy must be preceded or followed by a pulse of positive energy. These pulses must be close enough together so that any exotic effects due to the negative energy pulse can be compensated for by a following positive energy pulse. For example if a negative energy pulse fell into a black hole the energy of the black hole would decrease. This would also cause the entropy of the black hole to decrease [6]. This would violate the second law of thermodynamics which states that entropy cannot decrease. However if a positive energy pulse immediately followed the negative energy pulse then the entropy violation would only occur for a short period of time consistent with the uncertainty principle. However in the analysis in this section we have shown that a pulse of negative energy can exist and not be associated with a positive energy pulse. Therefore the quantum interest conjecture fails.

## 6. Temporal quantum inequality.

Due to the fact that the scalar potential is zero for $t \geq 0$ we expect that the temporal quantum inequality should apply for a sampling function that is non-zero only for $t \geq 0$. Let the sampling function $\rho_1(t)$ be defined by,

$$\rho_1(t) = \begin{cases} Nt^2(t-\tau)^2 & \text{for } \tau \geq t \geq 0 \\ 0 & \text{for } t < 0 \text{ or } t > \tau \end{cases} \tag{6.1}$$

where $N = 30/\tau^5$ and is chosen so that $\rho_1(t)$ satisfies the normalization condition. Using (6.1) in (1.4) we obtain,

$$\xi_{T,\min}[\rho_1] = -\frac{5}{3\pi\tau^2} \tag{6.2}$$

Next consider the energy density for $t \geq 0$ at some fixed point $x = \tau/2$ where we pick $\tau$ such that $\tau \gg a$. From (4.21) we obtain,

$$\Delta T_{00,\lambda}^{(+)}(\tau/2,t) = \frac{-\eta}{2}\left\{\left[\theta(\tau/2-t+a/2)-\theta(\tau/2-t-a/2)\right]\right\} \tag{6.3}$$

For the temporal quantum inequality to be obeyed the following expression must hold,

$$\int_0^\tau \Delta T_{00,\lambda}^{(+)}(\tau/2,t)\rho_1(t)dt \geq -\frac{5}{3\pi\tau^2} \tag{6.4}$$

Using (6.3) in the above and the fact that $\tau \gg a$ we obtain,

$$\frac{-\eta}{2}\int_{(\tau-a)/2}^{(\tau+a)/2}\rho_1(t)dt \geq -\frac{5}{3\pi\tau^2} \tag{6.5}$$

The left hand side of this expression can be integrated out to,

$$\frac{-\eta N}{2}\left\{\frac{(\tau+a)^5-(\tau-a)^5}{5\cdot 2^5} - \frac{\tau\left[(\tau+a)^4-(\tau-a)^4\right]}{2\cdot 2^4} + \frac{\tau^2\left[(\tau+a)^3-(\tau-a)^3\right]}{3\cdot 2^3}\right\} \tag{6.6}$$

For $\tau \gg a$ we obtain,

$$\int_0^\tau \Delta T_{00,\lambda}^{(+)}(\tau/2,t)\rho_1(t)dt \underset{\tau \gg a}{\cong} \frac{-15\eta a}{16\tau} \tag{6.7}$$

Use this in (6.5) to obtain,

$$\frac{-15\eta a}{16\tau} \underset{\tau \gg a}{\geq} -\frac{5}{3\pi\tau^2} \tag{6.8}$$

This relationship will not be true for $\tau \to \infty$ therefore the temporal quantum inequality fails.

### 7. The total kinetic energy must be positive.

There is one potential problem with the solution given by Mamaev and Trunov[8] which will be addressed in this section. Consider the situation for $t \geq 0$ after the scalar potential has been removed. The energy density at a given point is either negative or zero. Therefore the total energy integrated over all space is negative. This cannot be correct because the total energy cannot be less than or equal to zero.

In order to resolve this problem let us examine how a system in its initial unperturbed vacuum state evolves in time under the action of a scalar potential. Assume at some initial time, $t_1$, the scalar potential is zero, the state vector is $|0\rangle$, and the field operator is given by the initial unperturbed field operator,

$$\hat{\varphi}_0(x,t) = \sum_{j,\omega} \left( \hat{a}_{j\omega} f_{0,j\omega}(x,t) + \hat{a}^*_{j\omega} f^*_{0,j\omega}(x,t) \right) \tag{7.1}$$

Next apply a scalar potential $V(x,t) = c(t)V(x)$ where,

$$c(t) = \begin{cases} 0 \text{ for } t \leq t_1 \\ (t-t_1)/(t_2-t_1) \text{ for } t_1 < t < t_2 \\ 1 \text{ for } t > t_2 \end{cases} \tag{7.2}$$

We can think of $c(t)$ as turning on the potential during the interval $t_1$ to $t_2$, after which the potential remains constant in time. The field operator obeys the equation,

$$\frac{\partial^2 \hat{\varphi}}{\partial t^2} - \frac{\partial^2 \hat{\varphi}}{\partial x^2} + c(t)V(x)\hat{\varphi} = 0 \tag{7.3}$$

subject to the initial conditions $\hat{\varphi}(x,t_1) = \hat{\varphi}_0(x,t_1)$ and $\partial \hat{\varphi}(x,t_1)/\partial t = \partial \hat{\varphi}_0(x,t_1)/\partial t$.

In general we cannot solve this equation for a time dependent potential. However there are many cases were we can solve this equation for a static potential. We have just considered such a case in Section 2. Assume the voltage is positive and confined to some finite region, say $-a/2 \leq x \leq +a/2$, and zero outside of this region. Let's assume that $\hat{\varphi}_S(x,t)$ is the solution for the static potential. When the voltage is applied, at $t = t_1$, a disturbance is created which moves away from the finite region at the speed of light. At $t_2$ the voltage reaches it static value $V(x)$. After a sufficiently long time the solution $\hat{\varphi}(x,t)$ will settle down and it will approach its final value $\hat{\varphi}_S(x,t)$ over some large region $-L \leq x \leq +L$. We assume that, for $(t_f - t_2) \to \infty$, the solution $\hat{\varphi}(x,t_f) \to \hat{\varphi}_S(x,t_f)$ over some region $-L \leq x \leq +L$ where $L \to \infty$ and gets larger as $t_f$ increases.

This resolves the problem posed at the beginning at this section. For the problem considered here the kinetic energy density is given by Eq. (4.8) for the very large region $-L \leq x \leq +L$ where $L \to \infty$ for sufficiently large $t_f - t_2$. When the kinetic energy density is integrated over this region the result will be negative. Outside of this region the kinetic energy density will be positive which will make the total integrated kinetic energy positive as required. We are justified in ignoring this part of the solution because it is at infinity and therefore doesn't affect the result of our integrations over the sampling functions.

**8. Discussion.**

As noted in the Introduction the claims that have been made in this paper contradict previously established proofs of the quantum inequalities. Therefore if this paper is correct these proofs must contain errors. In this section we will examine a paper by Ford and Roman [2] that contains a proof of the quantum inequalities. In the next section we will look at a paper by Flanagan [3].

First let's review the results of the discussion so far. In this paper we work in the Heisenberg picture. That is, the state vector is constant in time. The time dependence is reflected in the field operators. The method of regulation is mode regulation. Each mode is identified and tracked through time. The kinetic energy density of each mode with respect to the unperturbed mode is determined and known for all time. The total energy density is the sum of the energy density of each mode.

The initial mode solutions are taken from the paper by Mamaev and Trunov [8]. They have shown that the kinetic energy density is negative within the region $|x| < a/2$ and zero outside of this region. What happens when the scalar potential is abruptly removed at $t = 0$? As explained previously the total kinetic energy does not change, however the local kinetic energy density (which is now equivalent to the energy density) can and does change. This change in the kinetic energy density is not instantaneous and the "rearrangement" of the local kinetic energy density does not occur faster than the speed of light. Therefore the kinetic energy density is continuous with respect to the removal of the potential. The effect of the removal of the potential is that the negative energy in the region $|x| < a/2$ will "radiate" outward at the speed of light.

For comparison we will focus on the proof of the quantum inequalities by Ford and Roman [2]. In their paper they use the "free" field operator, that is, in the notation of this discussion, their field operator is given by $\hat{\varphi}_0(x,t)$ with the mode solutions $f_{0,j\omega}(x,t)$. They use this field operator to obtain an expression for the energy density operator. They then prove that the quantum inequalities must hold for any possible state vector $|\Omega\rangle$.

There is a significant difference in approach between Ref. [2] and this discussion. In this discussion we start out with the field operator $\hat{\varphi}_\lambda(x,t)$ with the mode solution given by $f_{\lambda,j\omega}(x,t)$. The reason for this is that this field operator is appropriate to the problem at hand which is a scalar field in the presence of the scalar potential given by (3.3). It is important to point out that there is nothing special about the free field operator $\hat{\varphi}_0(x,t)$ compared to $\hat{\varphi}_\lambda(x,t)$. The free field operator $\hat{\varphi}_0(x,t)$ is the proper operator to use if the scalar potential has been zero for a very long time, in which case the mode solutions are given by $f_{0,j\omega}(x,t)$. The field operator $\hat{\varphi}_\lambda(x,t)$ is the proper one to use if the scalar potential $V_\lambda(x)$ has been present for a very long time in which case the mode solutions are $f_{\lambda,j\omega}(x,t)$. The question arises as to how to proceed when the potential is removed so that the field is technically a "free" field in that no scalar potential is present. However even though the field is a "free" field the mode solutions are certainly not $f_{0,j\omega}(x,t)$. They have been shown to be given by $f^{(+)}_{\lambda,j\omega}(x,t)$. Therefore a possible difference in the results is that free field operator $\hat{\varphi}_0(x,t)$ is not appropriate to this problem. The proof of the quantum inequality obtained in [2] may only apply to the field operator $\hat{\varphi}_0(x,t)$. The question then arises will the proof of the quantum inequalities presented in [2] work for the field operator $\hat{\varphi}_\lambda(x,t)$?. That is, the proof in Ref. [2] may be dependent on one's choice of field operator and may not be general.

In addition, there may be another problem with [2]. In the analysis in [2] it is assumed that the energy spectrum of the state vector $|\Omega\rangle$ is cut off at some maximum value. (See discussion in Appendix B of [2]). So, technically, this proof does not apply to all state vectors $|\Omega\rangle$ but only to state vectors that meet this criterion.

## 9. Flanagan's proof.

In this section we will discuss a potential problem with Flanagan's proof [3] of the quantum inequalities for a massless scalar field in 1-1 dimensional space-time. In the following we will show that there may be a problem with Flanagan's proof due to the fact that it relies on point-split regularization. It will be shown that point-split regularization can lead to spurious results and is not reliable.

Here we will consider temporal point-splitting. That is, the point splitting will only involve the time dimension. In this case the kinetic energy density operator for a scalar field with mass $m$ is written as,

$$\hat{T}_{00}(x,t;\alpha,[\varphi]) = \frac{1}{2}\left(\frac{\partial\hat{\varphi}\left(x,t+\frac{\alpha}{2}\right)}{\partial t}\frac{\partial\hat{\varphi}\left(x,t-\frac{\alpha}{2}\right)}{\partial t} + \frac{\partial\hat{\varphi}\left(x,t+\frac{\alpha}{2}\right)}{\partial x}\frac{\partial\hat{\varphi}\left(x,t-\frac{\alpha}{2}\right)}{\partial x} + m^2\hat{\varphi}\left(x,t+\frac{\alpha}{2}\right)\hat{\varphi}\left(x,t-\frac{\alpha}{2}\right)\right) \quad (9.1)$$

where $\alpha \to 0$.

To illustrate what is wrong with point splitting consider a special case of a free field where the unperturbed field operator is given by,

$$\hat{\varphi}_0(x,t) = \sum_k \frac{1}{\sqrt{2\omega_k L}}\left(\hat{a}_k e^{+ikx}e^{-i\omega_k t} + \hat{a}_k^* e^{-ikx}e^{+i\omega_k t}\right) \quad (9.2)$$

where $L \to \infty$ is the one-dimensional integration volume and $\omega_k = \sqrt{k^2 + m^2}$. In this case the kinetic energy density operator $\hat{T}_{00}(x,t;\alpha,[\varphi_0])$ is equivalent to the energy density operator because the scalar potential is zero.

The Hamiltonian operator is given by integrating the energy density operator over all space which yields,

$$\hat{H}_0(\alpha) = \int_{-L/2}^{+L/2} dx\, \hat{T}_{00}(x,t;\alpha,[\varphi_0]) = \frac{1}{2}\sum_k \omega_k\left(\hat{a}_k\hat{a}_k^* e^{-i\omega_k\alpha} + \hat{a}_k^*\hat{a}_k e^{+i\omega_k\alpha}\right) \quad (9.3)$$

Take the normal order of the above to obtain,

$$:\hat{H}(\alpha): = \sum_k \hat{a}_k^*\hat{a}_k \omega_k \cos(\omega_k\alpha) \quad (9.4)$$

The energy of the vacuum state is $\langle 0|:\hat{H}(\alpha):|0\rangle = 0$. Since the vacuum state is the state of minimum energy we expect that the energy of any other state will be positive. However this is not the case for the point split Hamiltonian given by the above expression. For example consider the state $\hat{a}_q^*|0\rangle$. The energy of this state is,

$$E_q = \langle 0|\hat{a}_q:\hat{H}(\alpha):\hat{a}_q^*|0\rangle = \omega_q\cos(\omega_q\alpha)$$

Pick a state where $\omega_q = \pi/\alpha$. In this case $E_q = (\pi/\alpha)\cos(\pi) = -(\pi/\alpha)$. Therefore when temporal point splitting is used we can have states with less energy than the vacuum state. Therefore point split regularization can lead to non-physical results and cannot be trusted. An example of this will be shown in the following discussion.

**10. A practical example.**

We will work a practical problem to demonstrate how the use of point split regulation leads to incorrect results. We will examine the vacuum expectation value of the kinetic energy associated with a massive scalar field in the presence of a scalar potential with point-like support in 1-1 dimension space-time. The kinetic energy will be calculated using two different methods of regularization, first using mode regularization and then using temporal point split regularization. It will be shown that the two methods give different results.

Assume that the field operator $\hat{\varphi}_\lambda(x,t)$ satisfies the equation,

$$\frac{\partial^2 \hat{\varphi}_\lambda}{\partial t^2} - \frac{\partial^2 \hat{\varphi}_\lambda}{\partial x^2} + m^2 \hat{\varphi}_\lambda + 2\lambda\delta(x)\hat{\varphi}_\lambda = 0 \tag{10.1}$$

where $2\lambda\delta(x)$ is the scalar potential and $\lambda \geq 0$. This problem was originally examined in [11] and is also discussed in [9].

Assume the boundary conditions at $x = \pm L/2$ are given by $\hat{\varphi}_\lambda(\pm L/2, t) = 0$. The field operator is given by,

$$\hat{\varphi}_\lambda(x,t) = \sum_k \left( \hat{a}_k \varphi_{\lambda,k}(x,t) + \hat{a}_k^* \varphi_{\lambda,k}^*(x,t) \right) \tag{10.2}$$

where the modes $\varphi_{\lambda,k}(x,t)$ are given by,

$$\varphi_{\lambda,k}(x,t) = e^{-i\omega_k t} \chi_{\lambda,k}(x) \tag{10.3}$$

The $\chi_{\lambda,k}(x)$ satisfy,

$$-\omega_k^2 \chi_{\lambda,k} - \frac{\partial^2 \chi_{\lambda,k}}{\partial x^2} + m^2 \chi_{\lambda,k} = -2\lambda\delta(x)\chi_{\lambda,k} \tag{10.4}$$

with the boundary condition $\chi_{\lambda,k}(\pm L/2) = 0$. The singularity at $x = 0$ yields the following relationships,

$$\chi_{\lambda,k}(0^+) = \chi_{\lambda,k}(0^-); \quad \frac{\partial \chi_{\lambda,k}(0^+)}{\partial x} - \frac{\partial \chi_{\lambda,k}(0^-)}{\partial x} = 2\lambda \chi_{\lambda,k}(0) \tag{10.5}$$

There are both even and odd solutions. The odd solutions are of the form $\chi_{\lambda,k} \propto \sin(kx)$. These solutions to not concern us because they are not affected by the potential due to the fact that they are equal to zero at $x = 0$ and therefore do not contribute to the change in the energy. They are not considered in the rest of this discussion. The solutions that are relevant are given by,

$$\chi_{\lambda,k} = A_k \cos(k|x| + \delta_k) \tag{10.6}$$

along with $\omega_k = \sqrt{k^2 + m^2}$ where $A_k$ is a normalization constant and,

$$\delta_k = -\arctan\left(\frac{\lambda}{k}\right) \tag{10.7}$$

From the boundary conditions we obtain $\cos((kL/2) + \delta_k) = 0$ which yields,

$$k = k_0 - \frac{2\delta_k}{L} \text{ where } k_0 = \frac{2\pi}{L}\left(n + \frac{1}{2}\right) \text{ with } n = 0, 1, 2, \ldots \tag{10.8}$$

The normalization constant $A_k$ is given by solving the normalization condition,

$$1 = -i \int_{-(L/2)}^{+(L/2)} \left( \varphi_{\lambda,k} \frac{\partial \varphi_{\lambda,k}^*}{\partial t} - \varphi_{\lambda,k}^* \frac{\partial \varphi_{\lambda,k}}{\partial t} \right) dx = 2\omega_k A_k^2 \int_{-(L/2)}^{+(L/2)} \left[\cos(k|x| + \delta_k)\right]^2 dx \tag{10.9}$$

From this and (10.8) we obtain,

$$A_k^2 = \frac{1}{\omega_k L \left(1 - (\sin(2\delta_k)/Lk)\right)} \tag{10.10}$$

From the above we have the following useful relationships,

$$\sin\delta_k = \frac{-\lambda}{\sqrt{k^2 + \lambda^2}}; \quad \cos\delta_k = \frac{k}{\sqrt{k^2 + \lambda^2}}; \quad \sin(2\delta_k) = \frac{-2k\lambda}{k^2 + \lambda^2}; \quad \cos(2\delta_k) = \frac{\lambda^2 - k^2}{k^2 + \lambda^2} \tag{10.11}$$

For a given mode the energy density is given by,

$$\varepsilon_{\lambda,k}(x) = \frac{1}{2}\left( \left|\frac{\partial \varphi_{\lambda,k}}{\partial t}\right|^2 + \left|\frac{\partial \varphi_{\lambda,k}}{\partial x}\right|^2 + m^2 |\varphi_{\lambda,k}|^2 \right) \tag{10.12}$$

Use (10.3) and (10.6) in this to obtain,

$$\varepsilon_{\lambda,k}(x) = \frac{A_k^2}{2}\left(\omega_k^2 + m^2 \cos\left[2\left(k|x| + \delta_k\right)\right]\right) \tag{10.13}$$

The total kinetic energy of the mode is given by integrating this quantity to obtain,

$$E_{\lambda,k} = \int_{-(L/2)}^{+(L/2)} \varepsilon_{\lambda,k}(x)\,dx = \frac{A_k^2}{2}\left(\omega_k^2 L - m^2 \frac{\sin(2\delta_k)}{k}\right) \tag{10.14}$$

The total kinetic energy is given by summing up the kinetic energies of each mode,

$$E_{\lambda,T} = \sum_{n=0}^{\infty} E_{\lambda,k} \tag{10.15}$$

Recall that $k$ is dependent on the summation index $n$ per Eq. (10.8). The kinetic energy of the mode for the free field is given by setting $\lambda = 0$ in (10.14) to obtain,

$$E_{0,k_0} = \frac{\omega_{k_0}}{2} \tag{10.16}$$

The total kinetic energy of the free field is then,

$$E_{0,T} = \sum_{n=0}^{\infty} E_{0,k_0} \tag{10.17}$$

The change in the kinetic energy is then given by,

$$\Delta E_T = E_{\lambda,T} - E_{0,T} \tag{10.18}$$

There is the usual problem with evaluating this expression due to the fact that $E_{\lambda,T}$ and $E_{0,T}$ are both infinite. That is why some type of regularization is required.

First we will evaluate this expression by using "mode" regularization. That is, we rewrite it as,

$$\Delta E_T = \sum_{n=0}^{\infty} E_{\lambda,k} - \sum_{n=0}^{\infty} E_{0,k} = \sum_{n=0}^{\infty} \Delta E_{\lambda,k} \tag{10.19}$$

where $\Delta E_{\lambda,k} = E_{\lambda,k} - E_{0,k_0}$ is change in the kinetic energy of the $k-th$ mode. Therefore we are calculating the change in the kinetic energy in each mode and summing all the changes to get the total change. In the limit that $L \to \infty$ we can use $\sum_{n=0}^{\infty} \to \int_0^{\infty} \frac{L\,dk_0}{2\pi}$ to obtain,

$$\Delta E_T = \frac{L}{2\pi} \int_0^{\infty} \left(E_{\lambda,k} - E_{0,k_0}\right) dk_0 \tag{10.20}$$

In the limit that $L \to \infty$ we can write,

$$A_k^2 \underset{L \to \infty}{=} \frac{1}{\omega_k L}\left(1+\left(\sin(2\delta_k)/Lk\right)\right)+O\left(1/L^2\right) \qquad (10.21)$$

where $O\left(1/L^2\right)$ means terms to the order $1/L^2$ or higher. Use this along with (10.11) in (10.14) to obtain,

$$E_{\lambda,k} = \frac{1}{2}\left(\omega_k - \frac{2\lambda k^2}{\omega_k L\left(k^2+\lambda^2\right)}+O\left(1/L^2\right)\right) \qquad (10.22)$$

Also we can obtain,

$$\omega_k = \sqrt{\left(k_0 - \frac{2\delta_k}{L}\right)^2 + m^2} \underset{L \to \infty}{=} \omega_{k_0} - \frac{2\delta_k k_0}{\omega_{k_0} L} + O\left(1/L^2\right) \qquad (10.23)$$

Use the above relationships to yield,

$$E_{\lambda,k} - E_{0,k_0} \underset{L \to \infty}{=} \frac{1}{2}\left(\frac{-2\delta_k k_0}{\omega_k L} - \frac{2\lambda k^2}{\omega_k L\left(k^2+\lambda^2\right)}+O\left(1/L^2\right)\right) \qquad (10.24)$$

Use this in (10.20) to obtain,

$$\Delta E_T = \frac{1}{4\pi}\int_0^\infty \left(\frac{-2\delta_k k_0}{\omega_k} - \frac{2\lambda k^2}{\omega_k\left(k^2+\lambda^2\right)}\right) dk_0 \qquad (10.25)$$

where the $O\left(1/L^2\right)$ term has been dropped. Use $k = k_0 + O\left(1/L\right) \underset{L \to \infty}{\cong} k_0$. Therefore we can replace $k$ with $k_0$ in the above expression to obtain,

$$\Delta E_T = -\frac{1}{2\pi}\int_0^\infty \left(\frac{\delta_{k_0} k_0}{\omega_{k_0}} + \frac{\lambda k_0^2}{\omega_{k_0}\left(k_0^2+\lambda^2\right)}\right) dk_0 \qquad (10.26)$$

This can be rewritten as,

$$\Delta E_T \underset{\Lambda \to \infty}{=} -\frac{1}{2\pi}\int_0^\Lambda \left(\frac{\delta_{k_0} k_0}{\omega_{k_0}} + \frac{\lambda k_0^2}{\omega_{k_0}\left(k_0^2+\lambda^2\right)}\right) dk_0 \qquad (10.27)$$

We will integrate the above expression as follows. First use (10.7) and integrate by parts to obtain,

$$\int_0^\Lambda \frac{\delta_{k_0} k_0}{\omega_{k_0}} dk_0 = -\int_0^\Lambda \frac{k_0 \arctan(\lambda/k_0)}{\omega_{k_0}} dk_0 = -\left(\left.\omega_{k_0} \arctan(\lambda/k_0)\right|_0^\Lambda + \lambda \int_0^\Lambda \frac{\omega_{k_0}}{(k_0^2 + \lambda^2)} dk_0\right) \quad (10.28)$$

Evaluate the first expression on the right,

$$\left.\omega_{k_0} \arctan(\lambda/k_0)\right|_0^\Lambda = \sqrt{\Lambda^2 + m^2} \arctan(\lambda/\Lambda) - (m\pi/2) \underset{\Lambda \to \infty}{=} \lambda - (m\pi/2) \quad (10.29)$$

Use this in (10.27) to obtain,

$$\Delta E_T = \frac{\lambda}{2\pi} - \frac{m}{4} + \frac{\lambda}{2\pi} \int_0^\infty \left(\frac{m^2}{\omega_{k_0}(k_0^2 + \lambda^2)}\right) dk_0 \quad (10.30)$$

Integrating out the last expression yields,

$$\Delta E_T = \frac{\lambda}{2\pi} - \frac{m}{4} + \left(\frac{m^2}{2\pi}\right) \begin{cases} \dfrac{1}{2\sqrt{\lambda^2 - m^2}} Log\left(\dfrac{\lambda + \sqrt{\lambda^2 - m^2}}{\lambda - \sqrt{\lambda^2 - m^2}}\right); \lambda > m \\ \dfrac{1}{\sqrt{m^2 - \lambda^2}} arctg\left(\dfrac{\sqrt{m^2 - \lambda^2}}{\lambda}\right); m < \lambda \end{cases} \quad (10.31)$$

In the limit that $\lambda \to \infty$ we obtain,

$$\Delta E_T \underset{\lambda \to \infty}{=} \left(\frac{\lambda}{2\pi}\right) - \left(\frac{m}{4}\right) \quad (10.32)$$

## 11. Regularization by Temporal point splitting.

In the last section the kinetic energy, $\Delta E_T$, was determined by mode regularization. In this section we will calculate the same quantity using temporal point split regularization. In this case the kinetic energy density of a given mode is given by,

$$\varepsilon_{\lambda,k}(x,t;\alpha) = \frac{1}{2}\left(\begin{array}{l} \dfrac{\partial \varphi_{\lambda,k}(x,t+\alpha/2)}{\partial t} \dfrac{\partial \varphi^*_{\lambda,k}(x,t-\alpha/2)}{\partial t} + \dfrac{\partial \varphi_{\lambda,k}(x,t+\alpha/2)}{\partial x} \dfrac{\partial \varphi^*_{\lambda,k}(x,t-\alpha/2)}{\partial x} \\ + m^2 \varphi_{\lambda,k}(x,t+\alpha/2) \varphi^*_{\lambda,k}(x,t-\alpha/2) \end{array}\right)$$

(11.1)

Using the analysis which led up to (10.14) the kinetic energy of this mode is given by,

$$E_{\lambda,k}(\alpha) = \frac{A_k^2}{2}\left(\omega_k^2 L - m^2 \frac{\sin(2\delta_k)}{k}\right) e^{-i\omega_k \alpha} \quad (11.2)$$

From (10.22) this becomes, for $L \to \infty$,

$$E_{\lambda,k}(\alpha) = \frac{1}{2}\left(\omega_k - \frac{2\lambda k^2}{\omega_k L(k^2+\lambda^2)} + O(1/L^2)\right)e^{-i\omega_k \alpha} \qquad (11.3)$$

The energy of the mode when $\lambda = 0$ is obtained by setting $\lambda = 0$ in (11.2) to yield,

$$E_{0,k_0}(\alpha) = \frac{\omega_{k_0}}{2}e^{-i\omega_{k_0}\alpha} \qquad (11.4)$$

Use (10.23) to obtain,

$$e^{-i\omega_k \alpha} = e^{-i\omega_{k_0}\alpha}\left(1 + \frac{2ik_0\delta_{k_0}\alpha}{\omega_{k_0}L}\right) + O(1/L^2) \qquad (11.5)$$

Using this and (10.23) in (11.3) to obtain

$$E_{\lambda,k}(\alpha) = \left(\frac{\omega_{k_0}}{2} + \frac{ik_0\delta_{k_0}\alpha}{L} - \frac{k_0\delta_{k_0}}{\omega_{k_0}L} - \frac{\lambda k^2}{\omega_k L(k^2+\lambda^2)} + O(1/L^2)\right)e^{-i\omega_{k_0}\alpha} + O(1/L^2) \qquad (11.6)$$

From this the total kinetic energy is,

$$E_{\lambda,T}(\alpha) = \frac{L}{2\pi}\int_0^\infty E_{\lambda,k}(\alpha)dk_0 = E_{\lambda,TA}(\alpha) + E_{\lambda,TB}(\alpha) + E_{\lambda,TC}(\alpha) \qquad (11.7)$$

where the terms on the right are defined as follows,

$$E_{\lambda,TA}(\alpha) = \frac{L}{2\pi}\int_0^\infty \frac{\omega_{k_0}}{2}e^{-i\omega_{k_0}\alpha}dk_0 \qquad (11.8)$$

$$E_{\lambda,TB}(\alpha) = \frac{i}{2\pi}\int_0^\infty k_0\delta_{k_0}\alpha e^{-i\omega_{k_0}\alpha}dk_0 \qquad (11.9)$$

$$E_{\lambda,TC}(\alpha) = -\frac{1}{2\pi}\int_0^\infty\left(\frac{k_0\delta_{k_0}}{\omega_{k_0}} + \frac{\lambda k_0^2}{\omega_{k_0}(k_0^2+\lambda^2)}\right)e^{-i\omega_{k_0}\alpha}dk_0 \qquad (11.10)$$

Evaluate the above in this limit $\alpha \to 0$. First consider $E_{\lambda,TB}(\alpha)$. In the limit $\alpha \to 0$ the integrand approaches zero unless $k_0$ is large. In this case we can replace $\alpha e^{-i\omega_{k_0}\alpha} \to \alpha e^{-ik_0\alpha}$ and $\delta_{k_0} \to -\frac{\lambda}{k_0}$. Therefore we obtain,

$$E_{\lambda,TB}(\alpha)\underset{\alpha\to 0}{=} -\frac{i\lambda\alpha}{2\pi}\int_0^\infty e^{-ik_0\alpha}dk_0 = -\frac{\lambda}{2\pi} \qquad (11.11)$$

where we have used the general relationship $\int_0^\infty e^{-ik_0\alpha} dk_0 = -i/\alpha$ for $\alpha \neq 0$.

Next consider $E_{\lambda,TC}(\alpha)$. In the limit $\alpha \to 0$ we can replace $e^{-i\omega_{k_0}\alpha} \to 1$ unless $k_0$ is large. If $k_0$ is large than the integrand in (11.10) goes as $(1/k_0^3)$. Therefore, under these conditions in the limit $\alpha \to 0$, $e^{-i\omega_{k_0}\alpha}$ can be replaced by the number one in (11.10). This yields $E_{\lambda,TC}(\alpha) = \Delta E_T$ (see Eq.(10.26)). Use these results in (11.8) to obtain,

$$E_{\lambda,T}(\alpha) = E_{\lambda,TA}(\alpha) - \frac{\lambda}{2\pi} + \Delta E_T \tag{11.12}$$

The kinetic energy of the unperturbed state is $E_{0,T}(\alpha) = E_{\lambda,TA}(\alpha)$. Therefore the regularized kinetic energy as determined using temporal point splitting is,

$$\Delta E_T(\alpha) = E_{\lambda,T}(\alpha) - E_{0,T}(\alpha) = -\frac{\lambda}{2\pi} + \Delta E_T \tag{11.13}$$

In the limit that $\lambda \to \infty$ we can use (10.32) to obtain,

$$\Delta E_T(\alpha)\Big|_{\lambda \to \infty} = -\left(\frac{m}{4}\right) \tag{11.14}$$

Therefore kinetic energy, as calculated using temporal point split regularization, is negative whereas the kinetic energy from mode regularization is positive. This suggests that point split regularization is not a reliable form of regularization and the use of this form of regularization may produce false results.

## 12. Summary and Conclusion.

Mamaev and Trunov have found the exact solution for a massless scalar field in the presence of the scalar potential $V_\lambda(x)$, defined by (1.7), and determined the kinetic energy density. This energy density is negative in the region $|x| < a/2$ and zero outside of this region. Due to the presence of a nonzero scalar potential this solution cannot be used to test the quantum inequalities because these only apply to the case where the external potentials are zero. This problem is rectified by instantaneously removing the potential at $t = 0$. When this happens we have shown the kiniteic energy density is continuous across the boundary at $t = 0$ and is now the same as the energy density. For $t > 0$ the energy density consists of two negative energy pulses moving at the speed of light. One moves in the positive direction and one in the negative direction. Therefore the spatial and temporal quantum inequalities do not hold. In addition the quantum interest conjecture fails.

**Appendix.**

From Ref. [8] the mode solutions are,

$$\chi_{\lambda,1\omega}(x) = \begin{cases} A_{1,\omega}(\Lambda,\Omega)\sin(\omega x), & |x| < a/2 \\ \sin(\omega x + \delta_1 \varepsilon(x)), & |x| > a/2 \end{cases} ; \chi_{\lambda,2\omega}(x) = \begin{cases} A_{2,\omega}(\Lambda,\Omega)\cos(\omega x), & |x| < a/2 \\ \cos(\omega x + \delta_2 \varepsilon(x)), & |x| > a/2 \end{cases}$$

where $\Lambda = \lambda a/2$, $\Omega = \omega a/2$, and $\varepsilon(x) = +1$ for $x > 0$ and $\varepsilon(x) = -1$ for $x < 0$,

$$A_{1,\omega}^2 = \left[\sin^2\Omega + \left((\Lambda/\Omega)\sin\Omega + \cos\Omega\right)^2\right]^{-1} ; A_{2,\omega}^2 = \left[\cos^2\Omega + \left((\Lambda/\Omega)\cos\Omega - \sin\Omega\right)^2\right]^{-1}$$

$$\tan\delta_1 = \frac{-(\Lambda/\Omega)\sin^2\Omega}{\left(1 + (\Lambda/2\Omega)\sin(2\Omega)\right)} ; \tan\delta_2 = \frac{-(\Lambda/\Omega)\cos^2\Omega}{\left(1 - (\Lambda/2\Omega)\sin(2\Omega)\right)}$$

For the case where $\lambda = 0$ we obtain,

$$\chi_{0,1\omega}(x) = \sin(\omega x); \quad \chi_{0,2\omega}(x) = \cos(\omega x)$$

# References

1. H. Epstein, V. Glaser, and A. Jaffe, *Nonpositivity of the energy density in quantized field theories*, Nuovo Cim. **36** (1965) 1016-1022.

2. L.H. Ford and T.A. Roman, *Restrictions on Negative Energy Density in Flat Spacetime*, Phys. Rev. **D55** (1997) 2082-2089. (Also axXiv:gr-qc/9607003).

3. E.E. Flanagan, *Quantum inequalities in two dimensional Minkowski spacetime*, Phys. Rev. **D56** (1997) 4922-4926. (Also arXiv:gr-qc/9706006).

4. C.J. Fewster and S.P. Eveson, *Bounds on negative energy densities in flat spacetime,* Phys. Rev. **D58**, 084010, 1998. (Also arXiv:gr-qc/9805024).

5. L.H. Ford, *Negative Energy Densities in Quantum Field Theory,* arXiv:0911.3597.

6. L.H. Ford, *Quantum coherence effects and the second law of thermodynamics,* Proc. R. Soc. Lond. A **364** (1978) 227-236.

7. M. Morris, K. Thorne, and U. Yurtsever, *Wormholes, Time Machines, and the Weak Energy Condition*, Phys. Rev. Lett. **61** (1988) 1446-1449.

8. S.G. Mamaev and N.N. Trunov. *Vacuum Expectation Values of the Energy-Momentum Tensor of Quantized Fields on Manifolds of Different Topology and Geometry. IV.* Russian Physics Journal, Vol. 24, No. 2 (1981) 171-174.

9. V. M. Mostepanenko and N.N. Trunov. *The Casimir Effect and its Applications.* Oxford University Press, Oxford (1997).

10. L.H. Ford and T. A. Roman, *The quantum interest conjecture*, Phys. Rev. D60 (1999) 104018. axXiv:gr-qc/9901074.

11. S.G. Mamaev and N.N. Trunov. *Quantum effects in external fields determined by potentials with point like support.* Soviet J. Nucl. Phys. Vol. 35 (1982) 612-617.